\documentclass[aps,pra,twocolumn,shortbibliography]{revtex4-1}
\usepackage{graphicx,amsmath,amssymb,mathtools}
\usepackage{dcolumn,bm}
\usepackage{physics}
\usepackage{dsfont,xcolor}
\usepackage[capitalise]{cleveref}

\newcommand{\bk}{{\mathbf k}}
\newcommand{\bx}{{\bm x}}
\newcommand{\bxh}{\hat{{\bm x}}}
\newcommand{\xh}{\hat{{ x}}}
\newcommand{\Xh}{\hat{X}}
\newcommand{\Xih}{\hat{\bar{X}}}
\newcommand{\Ph}{\hat{P}}
\newcommand{\Qh}{\hat{Q}}
\newcommand{\Phs}{\hat{{\cal P}}}
\newcommand{\Qhs}{\hat{{\cal Q}}}
\newcommand{\id}{\mathds{1}}
\newcommand{\hh}{\hat{h}}
\newcommand{\hhs}{\hat{\mathfrak{h}}}
\newcommand{\vh}{\hat{v}}
\newcommand{\vhs}{\hat{\mathfrak{v}}}
\newcommand{\Uh}{\hat{U}}
\newcommand{\Uhs}{\hat{\cal U}}
\newcommand{\Ah}{\hat{A}} 
\newcommand{\Ahs}{\hat{\cal A}}
\newcommand{\Pih}{\hat{\bar{P}}}
\newcommand{\Pis}{\hat{\bar{ {\cal P}}}}
\newcommand{\Qih}{\hat{\bar{Q}}}
\newcommand{\Fh}{\hat{F}}
\newcommand{\Fhs}{\hat{\cal F}}
\newcommand{\Fih}{\hat{\bar{F}}}
\newcommand{\Fis}{\hat{\bar{\mathcal{ F}}}}
\newcommand{\Rh}{\hat{R}}

\begin{document}

\title{Local Topological Markers in Odd Dimensions}
\author{Joseph Sykes}
\author{Ryan Barnett}
\affiliation{Department of Mathematics, Imperial College London, London SW7 2AZ, United Kingdom}

\begin{abstract}
Local topological markers have proven to be a valuable tool for investigating systems with topologically non-trivial bands. 
Due to their local nature, such markers can treat translationally invariant systems and spatially inhomogeneous systems on
an equal footing. Among the most prevalent of these is the so-called Chern marker, which is available for systems in two spatial dimensions.
In this paper, we describe how to generalize this marker to 1d and 3d  systems, by showing that the relevant expressions
 accurately describe the phenomenon
of topological pumping given by the first and second Chern numbers in 1d and 3d respectively. In addition to providing general derivations,
we verify the markers by numerically considering  model Hamiltonians. These results will open the door for future studies including the
influence of disorder on topological pumping and topological phase transitions in odd-dimensional systems.
\end{abstract}

\maketitle

\section{Introduction}

Due to their potential applications, as well as fundamental importance, 
interest in topological quantum systems has remained high over recent years. 
Among the most prominent and basic effects such systems can exhibit  is that of topological pumping.
Topological pumping in one spatial dimension (1d) was originally theoretically described by Thouless where it was shown that
for gapped, adiabatically varied, non-interacting Fermionic systems, charge is pumped by integer amounts over each time period \cite{thouless1983quantization}. Furthermore, this integer value of 
charge is  given by the first Chern number of the time-dependent system. Effects directly related to such topological pumping have been realised experimentally in photonic systems \cite{kraus2012topological,verbin2015topological,wimmer2017experimental}, cold atomic gas systems \cite{lohse2016thouless,schweizer2016spin,nakajima2016topological,nakajima2020disorder}, as well as magneto-mechanical meta-materials and elastic lattice systems \cite{rosa2019edge,grinberg2020robust}. The development of the modern theory of polarization provided an alternative interpretation of this effect as a change of the macroscopic polarisation of the system \cite{resta1992theory,king1993theory,ortiz1994macroscopic,resta1998quantum,vanderbilt2018berry}.

In 3d systems a more recently understood effect involves the linear response of a system's magnetization when an external electric field is applied or, equivalently, the
linear response of the polarization when an external electric field is applied. Such a response is described by the
magnetoelectric polarizability. It was shown that the isotropic contribution to this polarizability is a purely geometric quantity which has coupling
strength given by \cite{qi2008topological,essin2009magnetoelectric,malashevich2010theory}
\begin{equation}
    \alpha_{\text{CS}} = \frac{e^2}{h}\frac{\theta_{\text{CS}}}{2\pi}.
\end{equation}
Above $\theta_{\text{CS}}$ is defined in terms of the following integral over the Brillouin zone 
\begin{equation}\label{CS}
    \theta_{\text{CS}} = -\frac{1}{4\pi} \int_{\rm BZ} d\mathbf{k}\; \epsilon_{ijl} \, \text{Tr} \Big( \Tilde{\mathcal{A}}_i \partial_j \Tilde{\mathcal{A}}_l - \frac{2i}{3} \Tilde{\mathcal{A}}_i \Tilde{\mathcal{A}}_j \Tilde{\mathcal{A}}_l \Big)
\end{equation}
and is often referred to as the Chern Simons axion coupling
where  $\epsilon_{ijl}$ is the Levi-Civita symbol and  $\Tilde{\cal A}_i$ is the Berry connection.

For time-periodic systems in  three spatial dimensions, 
it is known that the change of $\theta_{\text{CS}}$ over a full  period is related to the second Chern number of the system. As such, an analogous 3d topological pump would be characterised by the second Chern number. However, unlike the 1d pump where actual charge is pumped, it is less clear in general what physical quantity is pumped for the time-dependent 3D case in the absence of external magnetic and electric fields when $\theta_{\text{CS}}$ changes by an integer value over a pumping cycle. In \cite{taherinejad2015adiabatic}, it was shown
for a particular model
that it is the Berry curvature itself that  is pumped.

In parallel efforts, there has been considerable progress in theoretically understanding phenomena in 4d topological quantum systems
such as the quantum Hall effect 
\cite{kraus2013four,price2015four,ozawa2019topological}. The current experimental realisation of 4d topological systems (with two of the dimensions being `synthetic') has been achieved in both photonic and cold atomic gas systems \cite{zilberberg2018photonic,lohse2018exploring}.
Such 4d topological systems can be related to time-dependent 3d systems by the process of dimensional reduction \cite{qi2008topological}.

In a separate but related direction, another advance in the field of topological physics is that of
local topological markers. Generally speaking, such markers reduce to the system's topological number
when there is translational invariance, but do not rely on translational invariance for their
evaluation. Among the most prevalent of these topological markers is the so-called Chern marker \cite{bianco2011mapping}. Since the Chern marker does not require translational invariance, it has been used to probe topological phases and topological phase transitions of inhomogeneous systems \cite{Prodan2010Entanglement,tran2015topological,mitchell2018amorphous,Irsigler2019Interacting,Gerbert2020Local,ulcakar2020kibble,marrazzo2017locality,hayward2020effect}. Recent work analysed the non-equilibrium dynamics of the Chern marker for a system undergoing a quantum quench and found that topological currents arise within the system \cite{caio2019topological}. It was also shown that the Chern marker can be used to investigate topological aspects of non-Hermitian systems \cite{Song2019Non-Hermitian}. However, the current form of the Chern marker is restricted to systems in two spatial dimensions. 

In this paper we will develop local topological markers for both 1d and 3d time-dependant systems. 
These markers are related to the system's dipole moment in 1d and the  projected 
fully anti-symmetric moment in 3d, namely
 $\varepsilon_{ijl} \Ph \xh_i \Ph \xh_j \Ph \xh_l \Ph$ where $\Ph$ projects into occupied single-particle states. 
We show both analytically and numerically that the change in the 1d and 3d markers over a full time period is quantized to integer values given by the first and second Chern numbers of the time-dependent system. 
The topological pumping of such  systems is therefore directly reflected by the local markers and moreover, the 
3d marker can be related to the previously-discussed axion coupling.
We also provide a generalization of the Chern marker to 4d time-independent systems.

The paper is organized as follows. In \cref{section:Notation and even markers} the notation  is set, and basic quantities are introduced for later use. In this section we also provide some background on the well-studied 2d Chern marker and 
show how to generalize it to the
4d case.  \cref{Section: Topological pumping in one dimension} focuses on developing the 1d marker. Analytical
derivations are supplemented by comparison to numerical calculations for a particular time-dependent Hamiltonian.
 In  \cref{section: 3D topological pumping} we then move on to develop a topological marker suitable for three
 spatial dimensions. After considering previous relevant work, we provide a constructive derivation of the 3d marker
 using insights gained from the 1d case. This marker is also verified by numerically investigating a particular 3d model.
 Lastly, in  \cref{Section: Discussion and conclusion} we discuss possible future avenues of investigation for the
 odd-dimensional topological markers and conclude.

\section{Background, notation, and markers in even dimensions} \label{section:Notation and even markers}

In this section we introduce the notation that will be used throughout the paper. We also introduce the effective time-evolution operator and the operator Berry connection which are crucial in developing the topological markers for 1d and 3d systems. After reviewing the 2d Chern marker, 
we develop  a generalized Chern marker for the 4d case 
and show that it reduces to the second Chern number for translationally invariant systems.

\subsection{Notation and Band Structure Convention} \label{subsection: Notation}

In what follows, we will set the notation used in the remainder of the paper. In most of this work, we will
consider systems with open boundary conditions that are translationally invariant in the bulk. We will utilize the concept of `nearsightedness’ such that deep in the bulk, the density matrix can be approximated by the density matrix of an infinite translationally invariant system \cite{kohn1996density,prodan2005nearsightedness}. For the purpose of this paper the terms `translationally invariant' and `deep in the bulk' most often can be used 
interchangeably. 

We take the Hilbert space of the system to be spanned by the states $\ket{\bx}\otimes \ket{\alpha} $
where $\ket{\bx}$ denotes a state in a unit cell with lattice vector ${\bx}$ and
the states $\ket{\alpha} $ represent degrees of freedom within the unit cell. These states
are normalized as $\braket{\bx}{\bx'} = \delta_{\bx, \bx'}$ and $\braket{\alpha}{\alpha'} = \delta_{\alpha,\alpha'}$.
Such notation is particularly amenable to tight-binding models \cite{asboth2016short}.
Eigenstates of a Hamiltonian with discrete translational invariance, or Bloch states, can then be written as
    \begin{equation} \label{wave function decomp}
        \ket{\psi_{\mathbf{k}n}} = \ket{\mathbf{k}} \otimes \ket{u_{\mathbf{k}n}} \quad \text{where} \quad \Bigg\{
        \begin{aligned} &\braket{\mathbf{k}}{\mathbf{k}'} = \hspace{0.1em} \delta(\mathbf{k}-\mathbf{k}') \\ &\braket{u_{\mathbf{k}n}}{u_{\mathbf{k}n'}} = \delta_{n,n'}. \end{aligned}
    \end{equation}
Here, $\ket{u_{\mathbf{k}n}}$ is the cell-periodic portion of the Bloch state where $n$
denotes the band, $\bk$ is the wave vector,
and $\ket{\bk} = \frac{1}{\sqrt{V_{\rm BZ}}} \sum_{\bx} e^{i \bx \cdot \bk} \ket{\bx}$ where $V_{\rm BZ}$ is the volume of the 
Brillouin zone which is related to the unit cell volume by $V_c=(2\pi)^d/V_{\rm BZ}$ where $d$ is
the spatial dimension of the system. 
 
 We use the following convention for the position operator:
 \begin{equation}
 \bxh = \sum_{\bx} \bx \ket{\bx}\bra{\bx} \otimes \mathds{1},
 \end{equation}
 where the identity operator above acts on degrees of freedom within a unit cell. 
 Note that with this convention, $\bxh$ is insensitive to degrees of freedom within a unit cell.
 The components of $\bxh$ will be denoted by $\xh_i$ for $i=1,\ldots,d$.
 With this, $\bxh \ket{\bx} = \bx \ket{\bx}$.
The local unit-cell trace over internal degrees of freedom in unit cell at $\bx$ can be evaluated as
\begin{equation}
    {\rm tr}_\bx (\hat{O}) = \sum_\alpha   \big( \bra{\bx} \otimes \bra{\alpha}  \big) \hat{O} \big( \ket{\bx} \otimes \ket{\alpha} \big)
    \end{equation}
 for a general operator $\hat{O}$. The full trace of operator $\hat{O}$ can then be evaluated
 as ${\rm Tr}(\hat{O}) = \sum_\bx {\rm tr}_\bx (\hat{O})$. Note that, for a finite Hilbert space,
 the full trace will satisfy the cyclic property ${\rm Tr}(\hat{O}_1 \hat{O}_2)={\rm Tr}(\hat{O}_2 \hat{O}_1)$
 but the local trace generally will not.
 
The operator projecting into the occupied states, for a translationally invariant system, is given by
\begin{equation} \label{projector}
    \Ph = \sum_{n}^\text{occ} \int_{\rm BZ} d\bk \ket{\psi_{\mathbf{k}n}}\bra{\psi_{\mathbf{k}n}} = 
    \int_{\rm BZ} d{\bf k} \ket{\mathbf{k}}\bra{\mathbf{k}} \otimes \Phs_\mathbf{k},
\end{equation}
where $n$ runs over the occupied bands and $\Phs_\mathbf{k}$ is  given by
\begin{equation} \label{translationally invar P}
    \Phs_\mathbf{k} = \sum_n^\text{occ} \ket{u_{\mathbf{k}n}}\bra{u_{\mathbf{k}n}}.
\end{equation}
The wave-vector integrals like the above are to be taken over the Brillouin zone.
We will generally denote $\mathbf{k}$-dependent operators with scripted letters as was done here and sometimes drop the subscript 
$\mathbf{k}$ for brevity. Note that for a system without translational invariance, $\Ph$ (the projector into occupied states) can be constructed in a similar way, but $\Phs_\bk$ cannot. Finally we denote the
compliment of $\Ph$ as $\Qh= \mathds{1}- \Ph$, which projects into the unoccupied states of the system.

\subsection{Adiabatic Evolution} \label{subsection: Adiabatic evolution}

Throughout this paper, we restrict our focus to quantum dynamics confined to the adiabatic regime. 
Furthermore, we will only consider the case of time dependence that is periodic (with time period $T$).
To ensure adiabatic evolution, we take the dynamics of the system to be governed by an effective time-evolution operator described in this section which assures that if the system starts in the (instantaneous) ground state it will remain in the ground state. Choosing such an effective
evolution operator is a matter of practical convenience; the main results of the paper will continue to hold for the true evolution provided one is in the
adiabatic regime.

The effective adiabatic Hamiltonian is given by \cite{kato1950adiabatic,messiah1981quantum,kitaev2006anyons}
\begin{equation}
\label{adiabatich} 
    \hat{h} = i\big[\dot{\Ph},\Ph \big],
\end{equation}
where $\dot{\Ph} = \frac{d}{dt}\Ph$. Recall that $\Ph$ projects into a set of instantaneous eigenstates of
the original `parent' Hamiltonian.
From this, one can construct the adiabatic evolution operator by
solving the time evolution equation $i\frac{d}{dt} \hat{U} = \hh \Uh$ with $\Uh(t=0)=\id$. A straightforward calculation
shows that this evolution operator produces the correct adiabatic evolution of the projection operators:
$\Ph(t) = \Uh \Pih \Uh^\dagger$, $\Qh(t) = \Uh \Qih \Uh^\dagger$ where the operators with overhead bars
are to be evaluated at $t=0$. Furthermore, this adiabatic evolution can be seen to retain the correct
Berry phase information of the original time-dependent Hamiltonian. A crucial advantage of
this procedure is that adiabatic evolution is automatically ensured, no matter how
fast $\Ph$ changes with respect to time.

All of the previous results of this subsection hold in general. On the
other hand, for a translationally invariant system, one can simplify further to
\begin{equation} 
\hh = \int_{\rm BZ} d \bk \ket{\bk } \bra{\bk} \otimes \hhs_\bk
\end{equation}
where the effective Bloch Hamiltonian is $\hhs_\bk = i \big[\dot{\Phs}_\bk,\Phs_\bk \big]$.
The time-evolution operator can similarly be written as
\begin{equation}
\Uh = \int_{\rm BZ} d \bk \ket{\bk } \bra{\bk} \otimes {\Uhs}_\bk.
\end{equation}
In this work, we will restrict to time dependence that does not break discrete translational invariance in the bulk. Therefore, 
the time dependence of $\Uh$ (for instance) will be given by that of $\Uhs_\bk$.

An explicit expression for the evolution operator is given by
\begin{align}
\label{evolution}
\Uhs_{\bk}(t) = \sum_n \ket{u_{\bk n}(t)} \bra{u_{\bk n}(0)}
\end{align}
where the cell-periodic states $\ket{u_{\bk n}}$ can be used to construct the relevant projection operators
as $\Phs_{\bk}=\sum_n^{\rm occ} \ket{u_{\bk n}} \bra{u_{\bk n}}$ and
 $\Qhs_{\bk}=\sum_n^{\rm unocc} \ket{u_{\bk n}} \bra{u_{\bk n}}$.  
 There is a slight subtlety with \cref{evolution}:
 the gauge of the cell-periodic states appearing in it are fixed. In fact, these states do 
not in general correspond to eigenstates of the Hamiltonian of the original system.
 A discussion of these technicalities is given in Appendix \ref{aevolution}.

\subsection{Berry Connection and Curvature Operators} \label{subsection: Berry connection}

We now use the time-evolution operator established above to
define the operator Berry connection and highlight how this form is related to the conventional Berry connection. In what immediately follows, we consider a translationally invariant system.
We define the operator Berry connection as
\begin{equation} \label{time difference Berry connection}
    \Ah_i = \int_{\rm BZ} d\bk \ket{\mathbf{k}}\bra{\mathbf{k}} \otimes \Ahs_i \quad \text{with} \quad \Ahs_i = i \hspace{0.1em} \Pis \Uhs^\dagger \partial_{i} \Uhs \Pis
\end{equation}
where, as before, overhead bars indicate evaluation at $t=0$, and $\partial_i = \frac{\partial}{\partial k_i}$. This expression
has similarities with the connection used in Ref.\ \cite{avron1989chern}.
From the name of $\Ahs_i$ there is the insinuation that it is related to the conventional Berry connection  matrix $\Tilde{\mathcal{A}}_i$ which has elements $(\Tilde{\mathcal{A}}_i)_{nm} = i\braket{u_{\mathbf{k}n}}{\partial_i u_{\mathbf{k}m}}$ where $n$ and $m$ label occupied bands. Indeed, by
inserting $\mathcal{U}_{\bk} = \sum_m \ket{u_{\mathbf{k}m}(t)}\bra{u_{\mathbf{k}m}(0)}$ into
\cref{time difference Berry connection}
a direct calculation shows that 
\begin{equation} \label{A is change of Berry over time}
    \Ahs_i = \sum_{nm}^\text{occ} \ket{u_{\mathbf{k}n}(0)}\bra{u_{\mathbf{k}m}(0)} \big( \Tilde{\mathcal{A}}_i (t) - \Tilde{\mathcal{A}}_i (0) \big)_{nm}.
\end{equation}
As such we see that $\Ahs_i$ gives the change of the Berry connection over time.

As shown in the following section, see \cref{opA1}, an expression for the operator Berry
connection operator that does not rely on translational invariance is
\begin{equation} \label{opA}
\hat{A}_i = \Uh^\dagger \Ph \xh_i \Ph \Uh  -  \Pih  \xh_i  \Pih.
\end{equation}
This expression reduces to
\cref{time difference Berry connection} when translational invariance is present and will be used
extensively when we construct the topological marker for the 3d case.

The concomitant Berry curvature operator is defined as
\begin{equation}
\Fh_{ij} = -i [ \Xh_i, \Xh_j]
\end{equation}
where for convenience we have defined the projected position operators as
\begin{equation}
\Xh_i = \Ph \xh_i \Ph.
\end{equation}
For a translationally invariant system, using manipulations identical to those appearing later in \cref{subsection: CM in even dims}, the Berry curvature operator can be written as
\begin{equation}
\Fh_{ij} =   \int_{\rm BZ} d \bk  \; \ket{\bk} \bra{\bk} \otimes \Fhs_{ij} 
\end{equation}
where $\Fhs_{ij} = i \Phs \big[\partial_{i}\Phs,\partial_{j}\Phs  \big] \Phs$. 
A direct calculation shows that this is related to the conventional Berry curvature
$\tilde{\mathcal{F}}_{ij} = \partial_i \tilde{\mathcal{A}}_j -\partial_j \tilde{\mathcal{A}}_i - i [\tilde{\mathcal{A}}_i, \tilde{\mathcal{A}}_j]$ as 
$\Fhs_{ij} = \sum_{nm}^{\rm occ} \ket{u_{\bk n}(t)} \bra{u_{\bk m}(t)} (\tilde{\mathcal{F}}_{ij})_{nm}$.

Finally, as shown in  Appendix \ref{Appendix:UFU derivation}, an elegant connection exists between
the quantities described in this and the previous subsection:
\begin{equation}
\Uhs^\dagger \Fhs_{ij} \Uhs = {\cal \Pih} (\partial_i \Ahs_j - \partial_j \Ahs_i -i [\Ahs_i,\Ahs_j])   
{\cal \Pih }
+ \Fis_{ij}.
\end{equation}

\subsection{Chern Marker in Even Dimensions} \label{subsection: CM in even dims}

We are now in a position to discuss the Chern marker in even dimensions. In particular, we
will provide some background on (and a short derivation of) the Chern marker in 2d 
\cite{kitaev2006anyons,bianco2011mapping}. We also provide a generalized Chern marker
for the 4d case, which reduces to the second Chern number for a translationally invariant system.

We will make regular use of the following identity which 
is  derived in Appendix \ref{Appendix:PxQ derivation} and holds for translationally invariant
systems:
\begin{equation} \label{PxQ}
        \Ph \hat{x}_i \Qh = \int_{\rm BZ} d\bk \ket{\mathbf{k}}\bra{\mathbf{k}} \otimes (-i)\Phs_{\bk}\partial_{i}\Phs_{\bk}.
\end{equation}
The utility of this expression (valid for arbitrary dimensions) is quickly appreciated as it provides a direct verification (and derivation) of the Chern marker as we demonstrate in the following. The Chern marker for a 2d system is defined in terms of the unit cell trace as \cite{bianco2011mapping}
\begin{equation} \label{CM}
\begin{aligned}
    \mathcal{M}_{2}(\bx) &=  -\frac{2\pi i}{V_c} {\rm tr}_\bx \left( \big[ \Ph \hat{x} \Ph,\Ph \hat{y} \Ph \big] \right)  \\
    &=-\frac{4\pi}{V_c} \text{Im} \; {\rm tr}_\bx(\Ph \hat{x} \Qh \hat{y} \Ph).
\end{aligned}
\end{equation}
Now taking the system to be translationally invariant and inserting \cref{PxQ} and its Hermitian
conjugate for the expressions $\Ph \xh \Qh$ and $\Qh \hat{y} \Ph$ into the second line of 
\cref{CM}, 
one immediately finds that the Chern marker reduces to the first Chern number $C_1$ of the system \cite{thouless1982quantized}:
\begin{equation}
\begin{aligned}
    \mathcal{M}_{2}(\bx) &= \frac{i}{2\pi}  \int_{\rm BZ} d \bk  \; \text{Tr}\Big( \Phs \big[\partial_{k_x}\Phs,\partial_{k_y}\Phs  \big] \Phs \Big) \\
    &= \frac{i}{2\pi} \sum_n^{\rm occ} \int_{\rm BZ} d\bk \;  \epsilon_{ij}  \braket{\partial_{i} u_{\bk n}}{\partial_j u_{\bk n}}
     = C_1.
    \end{aligned}
\end{equation}

We are now in a position to generalize the Chern marker to four dimensions.
We proceed in the spirit of the previous paragraph, by writing down an expression and showing that it reduces
to the second Chern number for translationally invariant systems. 
We take the four-dimensional
Chern marker to be
\begin{equation} \label{CM4}
\begin{aligned}
    \mathcal{M}_{4}(\bx) &=- \frac{2\pi^2}{V_c} \epsilon_{ijlm} \;
    {\rm tr}_\bx ( \Ph \xh_i \Ph \xh_j \Ph \xh_l \Ph \xh_m \Ph).
    \end{aligned}
    \end{equation}
Next, we replace the second and fourth projection operators in the 
above expression with $\Ph=\mathds{1}-\Qh$
and use the antisymmetry of the Levi Civita symbol to obtain
\begin{equation}
\begin{aligned}
    \mathcal{M}_{4}(\bx) &=- \frac{2\pi^2}{V_c} \epsilon_{ijlm}
    {\rm tr}_\bx ( \Ph \xh_i \Qh \xh_j \Ph \xh_l \Qh \xh_m \Ph).
\end{aligned}
\end{equation}
Taking the system to be translationally invariant, 
the relation \cref{PxQ} can then be used for the quantities $\Ph \xh_i \Qh$ and
 $\Qh \xh_i \Ph$.
Then carrying out the local trace, one arrives at
\begin{align}
 \notag
    \mathcal{M}_{4}(\bx) &= \frac{-\epsilon_{ijlm}}{8\pi^2} \int_{\rm BZ} d\bk  \text{Tr} 
    \Big( \Phs (\partial_{i}\Phs)(\partial_{j}\Phs)(\partial_{l}\Phs)(\partial_{m}\Phs) \Phs \Big) 
    \\ &= \frac{\epsilon_{ijlm}}{32\pi^2} \int_{\rm BZ} d\bk \; {\rm Tr} (\tilde{\mathcal{F}}_{ij} \tilde{\mathcal{F}}_{lm})=C_2.
    \label{M4}
\end{align}
The right hand side of the above equation is the second Chern number $C_2$ \cite{nakahara2003geometry,qi2008topological}. The
expression (\ref{CM4}) for the 4d Chern marker is a secondary result of the present work.
                
As stated earlier, the Chern marker for 2d systems has been used to investigate a wide range of inhomogeneous topological systems, both classical and quantum. Thus, the formulation of the Chern marker in 4d opens the possibility of investigating 4d inhomogeneous topological systems.
The theoretical investigation of 4d systems is complimented by the development of experimental techniques allowing experimental access to these systems via synthetic 
dimensions \cite{ozawa2019topological}.

\section{Topological Marker in One Dimension}\label{Section: Topological pumping in one dimension}

In what follows we consider pumping in time-dependent one-dimensional systems, with the goal of
developing a local topological marker which describes this effect.   Though we motivate this marker
with a concrete model, the results are general. For the concrete model, we take  the
time-dependent Aubry-Andr\'{e} Hamiltonian which is given by \cite{aubry1980analyticity,marra2015fractional}
\begin{equation}
\begin{aligned}
    \hat{H} = - J \sum_{n} & ( \ket{n}\bra{n+1}+ {\rm h.c.}) \\ &-\Delta \sum_{n}\text{cos}\big(2\pi \alpha n - \phi(t)\big) \ket{n}\bra{n}
\end{aligned}
\end{equation}
where   $n$ takes on integer values and labels the lattice sites.
The first term describes hopping between sites, while the second term gives a time-dependent
on-site energy shift. 
The time-dependence is incorporated through $\phi (t)= 2\pi t/T$ where $T$ is the period.  We will set $\alpha = 1/3$ which extends the unit cell to encompass three sites, and so the resulting system has three energy bands.
The lowest band has a Chern number of one:
\begin{align}
C_1 = \frac{i}{2\pi} \int_0^T\! dt \! \int_{\rm BZ} \!dk \; [\braket{\partial_t u_k}{\partial_k u_k}-\braket{\partial_k u_k}{\partial_t u_k}]=1
\end{align}
where $\ket{u_k}$ is the periodic part of the instantaneous Bloch eigenstate for the lowest band. 
We therefore expect one unit of charge to be pumped through a unit cell per cycle for this band.

\begin{figure}[t]
    \includegraphics[width=0.45\textwidth, trim = {0, 2cm, 0, 2cm}]{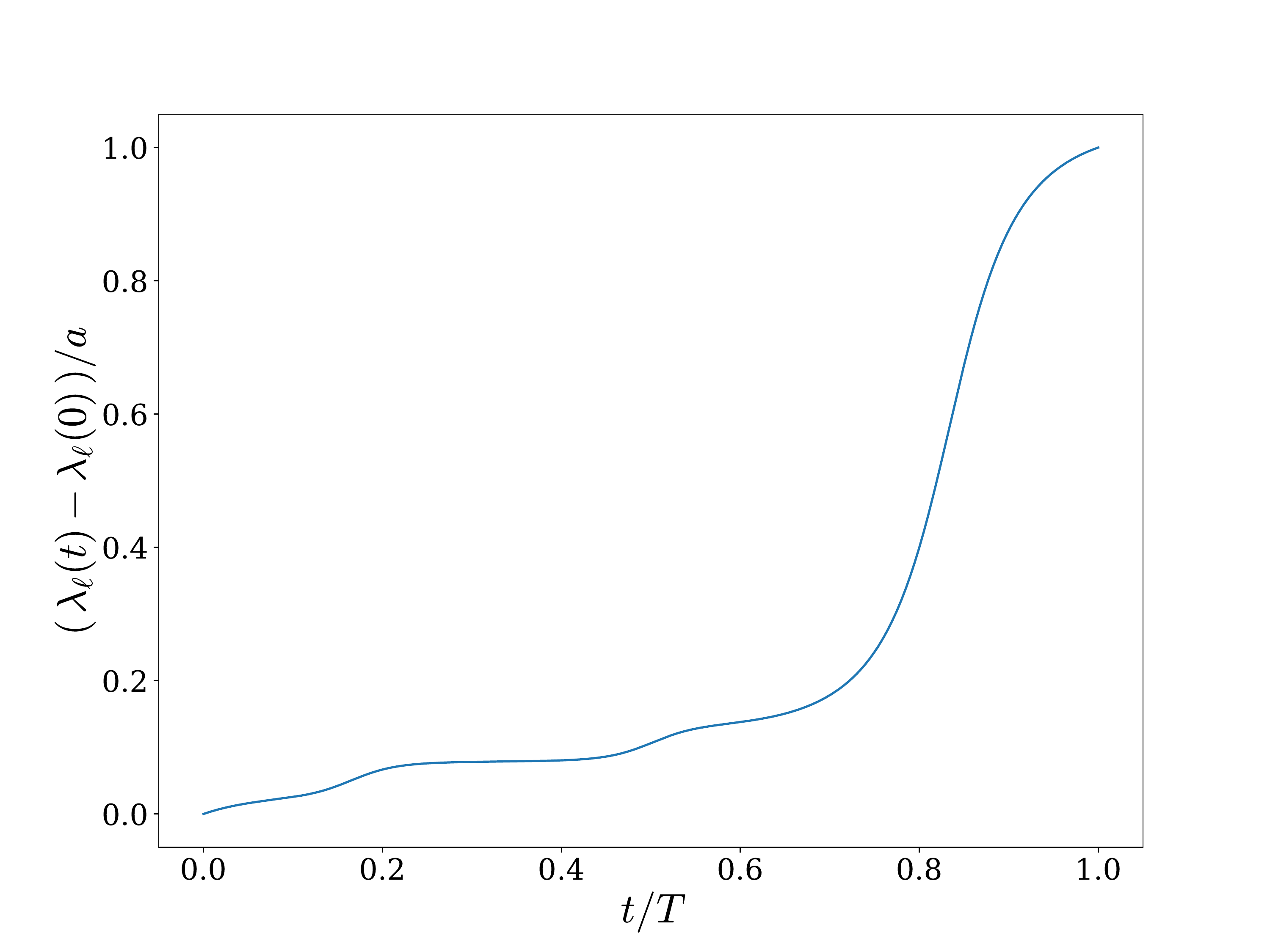}
    \caption{
    The position of a Wannier center $\lambda_\ell(t)$ localized deep in the bulk as a function of time.
    The lattice constant, comprising three lattice sites, is denoted by $a$. The parameters have been set to  $\Delta = 2J$, $\alpha= 1/3$, and the finite system is composed of 40 unit cells.}
    \label{fig:Wannier Centre}
\end{figure}

We now shift to considering a finite system with open boundary conditions.
We first consider eigenvalues of the operator $\Ph \xh \Ph$ 
where $\Ph$ projects into the lowest-energy `band'. That is, it projects 
out 2/3 of the total single-particle states which form the
higher-energy bands for the translationally invariant case.
\cref{fig:Wannier Centre} 
shows the position of one such eigenvalue deep within the bulk of the system
as a function of time. 
We refer to these eigenvalues, denoted by $\lambda_\ell$, as Wannier centers noting however that this term is often reserved for 
fully translationally invariant systems and we are analyzing a finite system that terminates.
One sees that the Wannier center increases by a lattice constant 
when evolved over a full period. 
It is well known from the modern theory of polarization \cite{vanderbilt2018berry}
that the change in the position of a Wannier center (in units of the lattice constant) is the same as the change of the polarization
$\Delta P_1$. Noting then that the change of the polarization is given by the integrated
current over a time period, $\Delta P_1 = \int_0^T dt J(t)$, we see that the behavior displayed
in \cref{fig:Wannier Centre}  describes the topological pump.

This might motivate one to consider the local trace of $\Ph \xh \Ph$ as a good potential candidate
for the 1d topological marker. However a short calculation shows that the local trace of this quantity deep
in the bulk is in fact time-independent. In particular, the calculation gives
\begin{equation} \label{PxP}
    \text{tr}_x (\Ph \xh \Ph) = \text{tr}_x (\xh \Ph - \Qh \xh \Ph) = x N_c;
\end{equation}
where we have used \cref{PxQ} to get the second equality and $N_c$ denotes the number
of particles per unit cell (which is one for this system). This shows that the local trace of the dipole operator is constant for all snapshots in time and thus the pumping is not reflected by this local quantity. 

With this shortcoming noted,
we will now consider a slight augmentation of the above and take
\begin{equation} \label{M1}
    {\cal M}_1(x ) = 
    \frac{1}{V_c}\text{tr}_\bx( \hspace{0.1em}\hspace{0.1em}  \Uh^\dagger \Ph \xh  \Ph \Uh  )
    =  \frac{1}{V_c}\text{tr}_\bx( \hspace{0.1em}\hspace{0.1em} \Pih \Uh^\dagger \xh \Uh \Pih ).
\end{equation}
It turns out that this expression
has all of the desired properties, which are elaborated upon below. We therefore identify it as the topological marker in 1d.
The pre-factor $1/V_c$, where $V_c$ is the unit cell volume (or lattice constant in 1d), is included to make the quantity
${\cal M}_1$
dimensionless.

We will now proceed to manipulate some of the terms appearing in \cref{M1}
to bring the expression into a more recognizable form.
For a translationally invariant system, using $\bra{k} \xh \ket{k'} = i  \partial_{k} \delta\big(k-k')$  and
$
\xh_j \ket{k} = -i \partial_j \ket{k}
$
it follows after an integration by parts that
\begin{equation} \label{UxU= x + UdU}
    \Uh^\dagger \xh \Uh = \xh + \int_{\rm BZ} d\bk \ket{\mathbf{k}}\bra{\mathbf{k}} \otimes i \Uhs^\dagger \partial_{k} \Uhs.
\end{equation}
Acting with projection operators and using \cref{time difference Berry connection} one finds
\begin{equation} \label{opA1}
\Pih \Uh^\dagger \xh \Uh \Pih = \Pih  \xh  \Pih  + \int dk \ket{k} \bra{k} \otimes \Ahs_k
\end{equation}
were \cref{time difference Berry connection} has been used. 
This establishes a direct relationship between the evolved position operator and the
operator Berry connection, similar to what appears in the semi-classical dynamics of wave-packets in
lattice systems \cite{xiao2010berry}.
In fact this equation gives a way
of defining the operator Berry connection in a way that does not depend on translational invariance:
$\Ah = \Uh^\dagger \Xh \Uh - \Xih$.

Now, 
for $x$ deep in the bulk and using the above relation, a direct calculation shows that
\begin{equation}
\begin{aligned}
   {\cal M}_1(x )  &= \frac{N_c}{V_c} x + \frac{1}{2\pi} \int_{\rm BZ} dk \; {\rm Tr} (\Ahs_k) \\
   &=  \frac{N_c}{V_c} x + \frac{1}{2\pi} \sum_n^{\rm occ}\int_{\rm BZ} dk \; \big(\tilde{{\cal A}}_k(t) - \tilde{{\cal A}}_k(0) \big)_{nn}
\end{aligned}
\end{equation}
where in the first equality we used \cref{time difference Berry connection}
and in the second equality, \cref{A is change of Berry over time}.  This can alternatively 
be written by using a fundamental relationship between Zak phase and polarization  \cite{king1993theory},
  as ${\cal M}_1 = \frac{N_c}{V_c} x+ \Delta P_1$.

We have thus shown that  ${\cal M}_1$ describes the topological pump, due to
its relationship with the polarization.
Due to the  local nature in real space of ${\cal M}_1$, one could use it to investigate time-dependent 1d systems with disorder or other spatial inhomogeneities. 
In hindsight, it is perhaps not very surprising that the local trace of $\Ph \xh \Ph$ does not
describe the polarization of the system since this quantity can be expressed purely in terms
of charge density at a particular time (and does not involve current) and it is known that the polarization cannot be expressed in terms of density alone. The time-evolution entering ${\cal M}_1$
gives the required history-dependence to determine the polarization.
As ${\cal M}_1(x)$ is a local quantity, there is a temptation to identify it with the
local polarization $P_1(x)$. 
For instance, with such an identification, and with a suitable continuum limit taken,
one could ask if the following basic relations to bound current and charge hold:
$\partial_t P_1(x) = J(x)$, $\partial_x P_1(x) = -\rho(x)$.
The veracity of such an identification is worth investigation, but
will not be considered further in this work.

\begin{figure}[t]
    \begin{center}
    \includegraphics[width=0.45\textwidth, trim = {0, 2cm, 0, 2cm}]{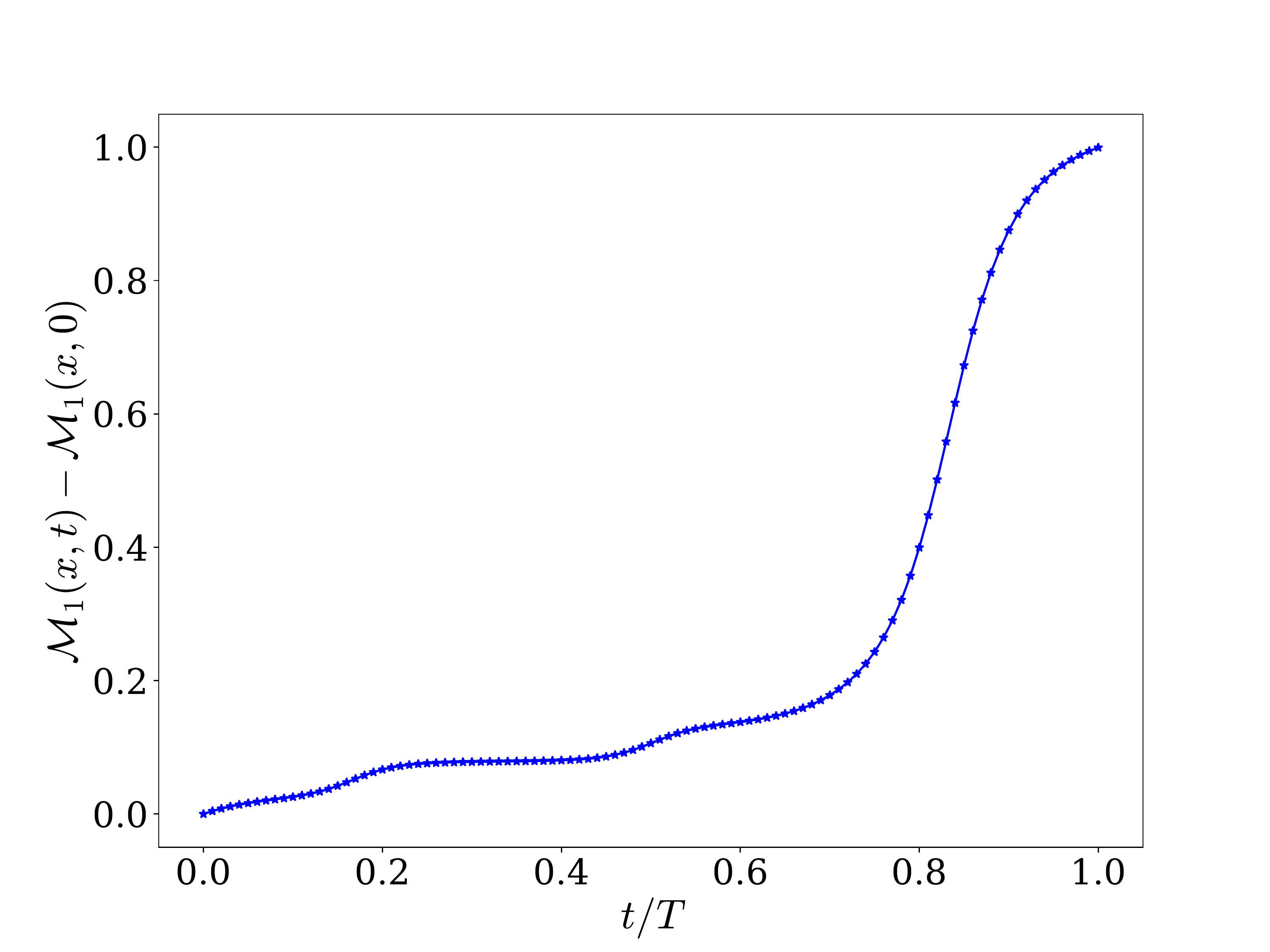}
    \end{center}
    \caption{ The change in the 1d marker ${\cal M}_1(x)$ over a full time period for the lowest band of the Aubry-Andre model with same parameters as \cref{fig:change in theta_1}. The local trace was taken over the 20th unit cell, $x = 20$, for a system size of 40 unit cells. }
    \label{fig:change in theta_1}
\end{figure}

With ${\cal M}_1$ defined we can now compute it numerically for the Aubry-Andr\'{e} model to verify the above considerations. \cref{fig:change in theta_1} shows  the time dependence of ${\cal M}_1$ for a finite system with open boundary conditions with 40 unit cells. The position of the unit cell for the local trace is taken
deep in the bulk of the system. Despite being a finite system, the result quantitatively 
matches that of the Wannier center plotted in  \cref{fig:Wannier Centre}. The result of 
\cref{fig:change in theta_1} was checked to be unchanged if the size of the total system
is increased.

The agreement discussed above becomes
worse when the local trace is taken over a unit cell close to the boundary of the system.
This is because the approximation of treating the system as translationally invariant breaks
down toward the edges. In fact, the sum over the local markers over the whole
system $\sum_{x} {\cal M}_1(x)$ can be seen to be periodic in time and so cannot
experience a net change over a pumping cycle. This is straightforwardly seen
by using the cyclic property of the trace, $\sum_{x} {\cal M}_1(x) =\frac{1}{V_c} {\rm Tr}( \Ph \xh \Ph)$,
and noting that $\Ph$ is periodic in time. This is closely related to the fact that  the
sum of the 2d Chern marker over a finite system vanishes: $\sum_{\bx} {\cal M}_2(\bx) =0$
which is again established by the cyclic property of the trace \cite{bianco2011mapping,caio2019topological}.

Before closing this section, 
we will briefly discuss how the numerical calculation of $\mathcal{M}_1(x)$
was done.
First, the time interval $[0,T]$ is discretized with sufficiently-small uniform spacing $\Delta t$.
For any discrete time point, $\Ph(t)$ can be calculated from the eigenvalues of $\hat{H}(t)$. 
Using $\Ph$ computed at $t+\Delta t$ and $t-\Delta t$,
$\dot{\Ph}(t)$ can be calculated using a finite difference method.
With these ingredients,  $\hat{h}$ can then be constructed using \cref{adiabatich}. Starting at $\Uh(0)=\id$, the time evolution operator at subsequent
times can be computed using $\Uh(t+\Delta t) = e^{-i \hat{h}(t)\Delta t} \Uh(t)$ which has an error of order
$(\Delta t)^2$, but retains the unitary form of $\Uh$ to numerical accuracy.
Lastly, $\mathcal{M}_1(x)$ is calculated using the second equality in \cref{M1}.
Convergence in the time spacing $\Delta t$, of course, should be checked.

To summarize, in this section we have introduced the 1d local topological marker ${\cal M}_1(x)$.  We
have demonstrated, both on general grounds and with a specific model, that it changes by an integer value
over a pumping cycle. This 1d marker satisfies a sum rule similar to that of the 2d Chern marker.
Though pumping in 1d systems is now well understood, 
the 1d marker allows systems with spatial 
inhomogeneities to be explored on an equal footing. 
Though some aspects in this section are well known, we presented them
as they provide 
important insights for the 3d case.

\section{Topological Marker in Three Dimensions} \label{section: 3D topological pumping}

We now move on to consider how to generalize to the case of three spatial dimensions. 
Like for the 1d case,  we will consider time-dependent Hamiltonians that do not
break discrete translational invariance in the
bulk.
We take an approach with several parallels
to the treatment in the previous section of the 1d case.

\subsection{Extensive Chern-Simons Axion Coupling} \label{subsection: theta_CS and sep Ham}

In previous leading works \cite{malashevich2010theory,olsen2017surface}, 
the following expression for the extensive Chern-Simons axion coupling was shown  to accurately
reflect the topological nature of 3d band structures:
\begin{equation} \label{slab theta_CS}
    \theta_{0} = \frac{4 i \pi^2 }{3} \epsilon_{ijl}\text{Tr} \big(\Ph \xh_i \Ph \xh_j \Ph \xh_l  \Ph \big)
\end{equation}
where the above trace is over the entire system.
In \cite{olsen2017surface}, a hybrid Wannier function basis was used to compute
the trace in the above expression. This basis is composed of states that are extended in the 
$x,y$-directions but localized in the $z$-direction. With such a basis, \cref{slab theta_CS} 
was written in terms of the hybrid Wannier centers.
Furthermore, pumping was demonstrated for the Wannier centers, and hence for \cref{slab theta_CS}, in a manner consistent with expectations from the second Chern number of the time-dependent system. For such a construction to work, the direction of pumping must be identified first. 
Then hybrid Wannier functions are chosen to be localized along this direction. If the ``wrong'' axis is taken for
the hybrid Wannier functions, then the resulting expression for \cref{slab theta_CS} would not capture the 
pumping.

 In a separate but related direction, the analysis of multipole moments, which do not involve projections between position operators, has proven very productive 
for analyzing the boundary of bulk insulators \cite{benalcazar2017quantized, benalcazar2017electric}. The connection between
the octupole moment and \cref{slab theta_CS} is one worth exploring, but will not
be pursued in the present work.

One would naively expect that a 3d local topological marker would have similarities
to \cref{slab theta_CS}.
In the following, we will demonstrate that a local expression following most directly from 
\cref{slab theta_CS} is unsuitable for our purposes. Replacing the full trace in 
\cref{slab theta_CS} with a unit-cell trace one obtains
\begin{equation} \label{theta_loc}
    \theta({\bx}) = \frac{4i\pi^2 }{3} \epsilon_{ijl} {\rm tr}_{\bx} \big(\Ph \xh_i \Ph \xh_j \Ph \xh_l \Ph \big).
\end{equation}
Now we restrict to positions $\bx$ deep within the bulk.
Working on a factor inside the trace, we have 
\begin{equation}
\epsilon_{ijl}\Ph \xh_j \Ph \xh_l \Ph = -\frac{1}{2} \epsilon_{ijl} \int_{\rm BZ} d\bk \ket{\bk} \bra{\bk}  \otimes  
\Phs [ \partial_j \Phs, \partial_l \Phs ] \Phs
\end{equation}
where \cref{PxQ} from \cref{subsection: CM in even dims} has been used. Next, using
$\Ph = \id - \Qh$ for the first projector appearing in \cref{theta_loc}, the local trace
can be carried out. One finds
\begin{equation} 
    \theta({\bx}) = - \frac{i V_c}{12 \pi} \epsilon_{ijl}  x_i 
    \int_{\rm BZ} d\bk  \;
{\rm Tr}\left( \Phs [ \partial_j \Phs, \partial_l \Phs ] \Phs \right).
\end{equation}
This expression for $\theta(\bx)$ depends on the Chern numbers of the various 2d planes of the system and the position of the unit cell over which the local trace is taken. This expression is 
in fact independent of time which can be seen as follows. Taking the time-derivative, we have
\begin{equation} \label{vanish}
    \frac{d \theta({\bx})}{dt} = - \frac{i V_c}{6 \pi} \epsilon_{ijl}  x_i 
    \int_{\rm BZ} d\bk  \; \partial_j
{\rm Tr}\left( \Phs [  \partial_t \Phs, \partial_l \Phs ] \Phs \right)=0.
\end{equation}
where we note that the quantity to the right of $\partial_j$ is a periodic function of $\bk$
and the cyclic property
of the trace has been used. 
Similar reasoning was used to show that Chern numbers are unaffected by a 
quantum quench \cite{caio2015quantum}.
In contrast to ${\cal M}_1(x)$ considered in \cref{Section: Topological pumping in one dimension},
$\theta(\bx)$ is  time-independent and so cannot describe pumping. In fact, the above
is very reminiscent of the fact that in 1d, ${\rm tr}_x (\Ph \xh \Ph)$ is time independent,
see \cref{PxP}.

Furthermore, it can be shown that $\theta_0$ will identically vanish for some finite systems exhibiting
topological pumping. We consider a Hamiltonian which decouples in the following way
\begin{equation} \label{decoupled H}
 \hat{H} = \hat{H}_1\otimes \mathds{1} + \mathds{1}\otimes \hat{H}_2.
\end{equation}
Hamiltonians that can be written in this way have been recently considered in the context of the 4d quantum Hall effect
\cite{kraus2013four, price2015four}.
The operators $\hat{x},\hat{y}$ act on the Hilbert space of $\hat{H}_1$ and the operator $\hat{z}$ acts on the Hilbert space of $\hat{H}_2$. The eigenfunctions of $\hat{H}$ all have the separable form
$
 \ket{\phi_l} \otimes \ket{\chi_m}
$
where $\ket{\phi_l}$ and $\ket{\chi_m}$ individually are eignenstates of $\hat{H}_1$ and $\hat{H}_2$
respectively.
Using this one can define the projector $\Ph$ as
\begin{equation} \label{decoupled projector}
    \Ph = \sum_{l,m} a_{lm} \ket{\phi_l}\bra{\phi_l} \otimes \ket{\chi_m}\bra{\chi_m},
\end{equation}
where $a_{lm}$ is either one or zero for occupied and unoccupied states respectively and the summation is over all values of $l$ and $m$.
Next we use the cyclic property of the trace to obtain
\begin{equation} \label{antisymetric trace of PxPyPz seperable}
    \epsilon_{ijl} \text{Tr} (\Ph \xh_i \Ph \xh_j \Ph \xh_l) = 3\big( \text{Tr}( \Ph \xh \Ph \hat{y} \Ph \hat{z} ) - \text{Tr} (\Ph \hat{y} \Ph \xh \Ph \hat{z} ) \big).
\end{equation}
Using \cref{decoupled projector} one can show that $\text{Tr}(\Ph \xh \Ph \hat{y} \Ph \hat{z})$ takes the form
\begin{equation}
\begin{aligned}
    \text{Tr}\big( \Ph \xh \Ph \hat{y} \Ph \hat{z} \big) = \sum_{l,l',m} a_{lm}a_{l'm}\bra{\phi_l} \xh & \ket{\phi_{l'}}\bra{\phi_{l'}} \hat{y} \ket{\phi_l} \\
    &\times \bra{\chi_m} \hat{z} \ket{\chi_m}
\end{aligned}
\end{equation}
where $(a_{lm})^2=a_{lm}$ has been used. Now one can observe that the left-hand side of
the above equation is unaffected if $\hat{x}$ and $\hat{y}$ are interchanged. This means that 
the two terms in \cref{antisymetric trace of PxPyPz seperable} cancel one another and so
the axion coupling vanishes identically $\theta_0=0$ for this case of a decoupled Hamiltonian.

It is interesting to contrast this observation with the closest analog of $\theta_0$ for the 1d
case, namely the system's dipole moment: $p(t)={\rm Tr} ( \Ph(t) \xh)$. 
Although $p(t)$ is necessarily periodic in time for the finite system 
(since $\Ph(t)$ is), pumping behaviour can be
deduced from it.
That is, for large but finite systems,
the time dependence of the dipole moment will be directly proportional to the position of a Wannier
centre deep in the bulk, see \cref{fig:Wannier Centre}, apart from abrupt jumps corresponding Wannier
centers exiting one side of the system and appearing on the other. Such jumps are very analogous to the
$2\pi$ phase ambiguity of the Zak phase, and serve to enforce the periodicity of $p(t)$. 
One might naively expect similar behavior to be encapsulated in
$\theta_0$. However,  for the case of a decoupled Hamiltonian as describe above, it is not.

\subsection{Derivation of the 3d Marker} \label{Subsection: 3d marker}
With the motivation outlined in the previous section, and with the shortcomings of a local version of
$\theta_0$ highlighted, we now deduce a suitable expression for the 
3d topological marker, ${\cal M}_3$.
Noting how time dependence was incorporated in the 1d case, e.g.\  \cref{M1}, and the axion coupling considered in the previous section,
we consider the following operator as a potentially fruitful starting point:
\begin{equation}
\Rh = \epsilon_{ijl} \Uh^\dagger  \Xh_i \Xh_j \Xh_l    \Uh
\end{equation}
where we recall that $\Xh_i = \Ph \xh_i \Ph$. We will proceed by differentiating this quantity
with respect to time and identifying which terms of 
$\int_0^{T} dt \; {\rm tr}_{\bx} \left( d \Rh / dt \right)$
are proportional to the second Chern number.

To facilitate this we first introduce a small amount of additional notation. The projected velocity
operator is 
\begin{equation}
\hat{v}_i = -i \Ph [\xh_i,\hat{h}] \Ph
\end{equation}
where $\hat{h}$ is the Hamiltonian governing the dynamics. For the effective adiabatic
Hamiltonian, $\hat{h} = i [\dot{\Ph} ,\Ph]$, this becomes $\hat{v}_i = \Ph \dot{\Xh}_i \Ph$.
For the case where translational invariance is present, the projected velocity operator
can be written as
\begin{equation}
\hat{v}_i = \int_{\rm BZ} d \bk \ket{\bk} \bra{\bk} \otimes \vhs_i
\end{equation}
where $\vhs_i = \Fhs_{ti} = i \Phs [\partial_t \Phs,\partial_i \Phs ]  \Phs $.
A short calculation shows that the operator Berry connection $\Ah_j = \Uh^\dagger \Xh_j \Uh - \Xih_j$
(overhead bars continue to denote evaluation at $t=0$) is related to the projected
velocity operator  as
\begin{equation}
\frac{d}{dt} \Ah_i = \Uh^\dagger \hat{v}_i \Uh.
\end{equation}

Now, returning to the operator $\Rh$, by using the relations from the previous paragraph, its
time derivative can be seen to be
\begin{equation}\label{dR/dt}
\begin{aligned}
\frac{d \hat{R}}{dt} &=  \epsilon_{ijl} 
\Uh^\dagger \left( \hat{v}_i \Xh_j \Xh_l + \Xh_i \hat{v}_j \Xh_l   + \Xh_i \Xh_j \hat{v}_l   \right) \Uh \\
&= \hat{L}_1 + \hat{L}_2 + \hat{L}_3
\end{aligned}
\end{equation}
where the  operators $\hat{L}_{i}$ are defined  by the three terms in the preceding expression respectively.
When taking a local trace of this expression deep within the bulk the first and last terms can be seen to give equal contributions. As well as this, when integrated over time, both terms give a contribution proportional to the second Chern number. For instance, for the translationally invariant case, the first term can be written as
\begin{equation}
\begin{aligned}
\hat{L}_1&=\epsilon_{ijl}  \Uh^\dagger  \hat{v}_i \Xh_j \Xh_l  \Uh = 
\frac{i}{2}   \epsilon_{ijl} \Uh^\dagger  \hat{v}_i \Fh_{jl} \Uh \\
&= \frac{i}{2}  \epsilon_{ijl} \int_{\rm BZ} d\bk \; \ket{\bk} \bra{\bk} \otimes \Uhs^\dagger \Fhs_{ti} \Fhs_{jl} \Uhs.
\end{aligned}
\end{equation}
The local trace evaluated deep in the bulk is then readily evaluated to be
\begin{equation}
\begin{aligned}
{\rm tr}_{\bx}(\hat{L}_1) & = \frac{i}{2 V_{\rm BZ}} \epsilon_{ijl} \int_{\rm BZ} d\bk \; {\rm Tr} (\Fhs_{ti} \Fhs_{jl}) \\ 
& = \frac{i}{8 V_{\rm BZ}} \epsilon_{\mu \nu \sigma \rho} \int_{\rm BZ} d\bk \; {\rm Tr} (\Fhs_{\mu \nu} \Fhs_{\sigma \rho})
\end{aligned}
\end{equation}
where the Greek indices are summed over both space and time components. In arriving at the last equality
the cyclic property of the trace, as well as the antisymmetry of $\Fh_{ij}$, was used.
We are using a convention such that $\epsilon_{t k_x k_y k_z} = -\epsilon_{ k_x k_y k_z t} =1 $.
Then, integrating over a time period, we have
\begin{align} \label{propto}
\int_0^T dt \; {\rm tr}_{\bx}(\hat{L}_1)  = \frac{4i \pi^2}{V_{\rm BZ}} C_2
\end{align}
where $C_2$ is the second Chern number. A very similar calculation gives the same result
for $\hat{L}_3$. Note that ${\rm tr}_{\bx} (\hat{L}_1) = {\rm tr}_{\bx} (\hat{L}_3)$ deep in the bulk.

We now are left to deal with the middle term in \cref{dR/dt}, $\hat{L}_2$, which does not have such a 
simple relation to the second Chern number. With a straightforward but tedious manipulation,
this contribution can be rewritten as
\begin{align} \label{tedious}
\hat{L}_2  \notag
=&  \epsilon_{ijl}  \left[ \frac{1}{2} \Xih_i  + \frac{1}{3} \hat{A}_i , \;\Uh^\dagger \vh_j \Uh \Ah_l  \right] 
- {\rm h.c.}\\
&+ \epsilon_{ijl}  \frac{d}{dt} \left(  \frac{1}{6} \Ah_i \Ah_j \Ah_l  
+ \frac{i}{4} \Xih_i  \Uh^\dagger \Fh_{jl} \Uh    
\right. \\ &  \;\; \;\;\;  - \left.  \frac{i}{4} \hat{A}_i   \Fih_{jl}    \right)
-  {\rm h.c.} \notag
\end{align}
where extensive use of the antisymmetry of $\epsilon_{ijl}$ has been used.
When taking the local trace deep in the bulk, the term involving the commutator will vanish.
Evaluating the local trace of the first contribution involving $\Xih_i$  leads to a Brillouin zone
integral of the derivative of a periodic quantity. See Appendix \ref{Appendix:PxQ derivation}
for more details.
The local trace from the second contribution involving $\Ah_i$ can be seen to vanish by
noting that each factor involved is translationally invariant in the bulk, and using the cyclic property of the trace.

With the considerations from the previous paragraph, we define the new operator
\begin{align}
\hat{R}' = \hat{R} + \frac{i\epsilon_{ijl} }{4}\left(\hat{A}_i   \Fih_{jl} 
- \Xih_i  \Uh^\dagger \Fh_{jl} \Uh
+\frac{2i}{3} \Ah_i \Ah_j \Ah_l  + {\rm h.c.}     \right).
\end{align}
The terms in parenthesis above follow directly from \cref{tedious}. We then have that
$\int_0^{T} dt \; {\rm tr}_{\bx} \left( d \Rh' / dt \right)$ is proportional to the second Chern number for
$\bx$ deep in the bulk.
The constant of proportionality can be read off from \cref{propto}.  This then gives our expression for
the 3d marker
\begin{align}\label{M3}
{\cal M}_3(\bx) \!= \!\frac{\pi \epsilon_{ijl} }{2V_c} {\rm Re} \! \left[{\rm tr}_{\bx} \!\! \left( \! \Ah_i \Uh^\dagger \Fh_{jl} \Uh 
\!+ \! \Ah_i \Fih_{jl} 
\!+ \! \frac{2i \Ah_i \Ah_j \Ah_l }{3}\right) \! \right]\!.
\end{align}
When $\bx$ is deep in the bulk we have that the change of ${\cal M}_3$ over a time cycle is the
second Chern number. It is also worth noting that none of the quantities in this expression rely upon
translational invariance for their evaluation. 
\cref{M3} is the main result of this paper.
For convenience we collect the relevant quantities 
entering \cref{M3} all together here: $\Ph$ projects into the occupied single-particle states, $\Ah_i = \hat{U}^\dagger \Ph \xh_i \Ph \Uh-\Pih \xh_i \Pih$, 
$\Fh_{ij}=-i [\Ph \xh_i \Ph, \Ph \xh_j \Ph]$,  $\Uh$ is the time evolution operator for 
$\hat{h} = i [\dot{\Ph}, \Ph]$, and overhead bars denote evaluation at time $t=0$.

\subsection{Relation Between the 3d Marker and Axion Coupling}

The behavior of ${\cal M}_3$ found in the
previous section, namely that it changes by the second Chern number over a time period,  suggests it has close connection with the Chern Simons
axion coupling parameter $\theta_{\rm CS}$, \cref{CS}.
It is interesting to note though that the 3d marker follows directly
from the occupied-states projection operator. That is, all constituent elements (e.g.\ 
$\Ah_i$, $\Fh_{ij}$, $\hat{U}$, $\hat{h}$)  follow from $\Ph$ alone. As such, ${\cal M}_3$
is manifestly gauge invariant. On the other hand, it is well known that the axion coupling
(which is the integrated Chern-Simons three-form) \cref{CS} is only gauge invariant modulo
$2\pi$ \cite{nakahara2003geometry}.

In order to evaluate the integral in \cref{CS}, one typically needs to be able to define
smooth and periodic states $\ket{u_{\bk n}} $ as a function of $\bk$ across the entire Brillouin zone. That is, it is typically assumed that there is
no topological obstruction which follows if all of the first Chern numbers of the system vanish. 
For a discussion of such subtleties, see \cite{moore2017introduction}. Accordingly, in this
section we will assume that there are no topological obstructions to defining smooth 
and periodic Bloch states across the Brillouin zone. Note, however, if one is working only with the
marker, such assumptions are not needed.

Deep in the bulk, the marker evaluated at time $t$ takes the form
\begin{align}\label{CS_derive}
{\cal M}_3 =\frac{\epsilon_{ijl}}{16 \pi^2} \int_{\rm BZ} d\bk  {\rm Tr} \left( \Ahs_i \Uhs^\dagger \Fhs_{jl} \Uhs + \Ahs_i \hat{\bar{{\cal F}}}_{jl} 
+ \frac{2i}{3} \Ahs_i \Ahs_j \Ahs_l \right).
\end{align}
 To connect this to the axion coupling, we can use the results in
 \cref{subsection: Berry connection}
to express the above in terms of $\tilde{{\cal A}}_i$ and $\tilde{{\cal F}}_{ij}$.
For instance,  in doing so one finds that the third term in the parenthesis
in \cref{CS_derive} is replaced with 
$\frac{2i}{3}(\tilde{\cal A}_i(t) -\tilde{\cal A}_i(0)) 
(\tilde{\cal A}_j(t) -\tilde{\cal A}_j(0))
(\tilde{\cal A}_l(t) -\tilde{\cal A}_l(0))
$
and ${\rm Tr}$ becomes the usual matrix trace.
Multiplying the terms out and using the cyclic property of the trace leads to
\begin{equation}
\begin{aligned}
{\cal M}_3 =&- \frac{1}{2\pi} (\theta_{\rm CS} (t) - \theta_{\rm CS}(0)) \\
&+\frac{\epsilon_{ijl}}{8 \pi^2} \int_{\rm BZ} d\bk \partial_i  {\rm Tr} [\tilde{\cal A}_j(0) \tilde{\cal A}_l(t)].
\end{aligned}
\end{equation}
where $\theta_{\rm CS}(t)$ is determined by evaluating \cref{CS} at time $t$.
From the assumption stated earlier, the final term vanishes and so we are left with 
simply
\begin{align}
{\cal M}_3 =&- \frac{1}{2\pi} (\theta_{\rm CS} (t) - \theta_{\rm CS}(0)).
\end{align}
Hence the 3d marker deep in the bulk at time $t$ corresponds to the change over time
of the integrated Chern Simons three form. This is similar to the fact that the 1d marker
 gives the change of polarization.
Note, however, that since the marker is gauge
invariant, and does not rely on translational invariance, it is perhaps the more fundamental quantity.

\subsection{3d Model} \label{Subsection: 3D topological model}
With the 3d marker now defined, we will proceed to verify its behavior with a specific model system.
We first construct a decoupled Hamiltonian, having the general form of \cref{decoupled},
 by combining the Aubry-Andre model and the Harper-Hofstadter model in a specific way \cite{hofstadter1976energy,aubry1980analyticity}. 
 Constructions in a similar spirit have been carried out to investigate the quantum Hall effect in 4d
 \cite{kraus2013four,price2015four}.
 We then calculate the 3d marker for the corresponding
 finite system with open boundary conditions.

\begin{figure}[t]
    \begin{center}
    \includegraphics[width=0.40\textwidth, trim = {0, 2cm ,0 ,2cm}]{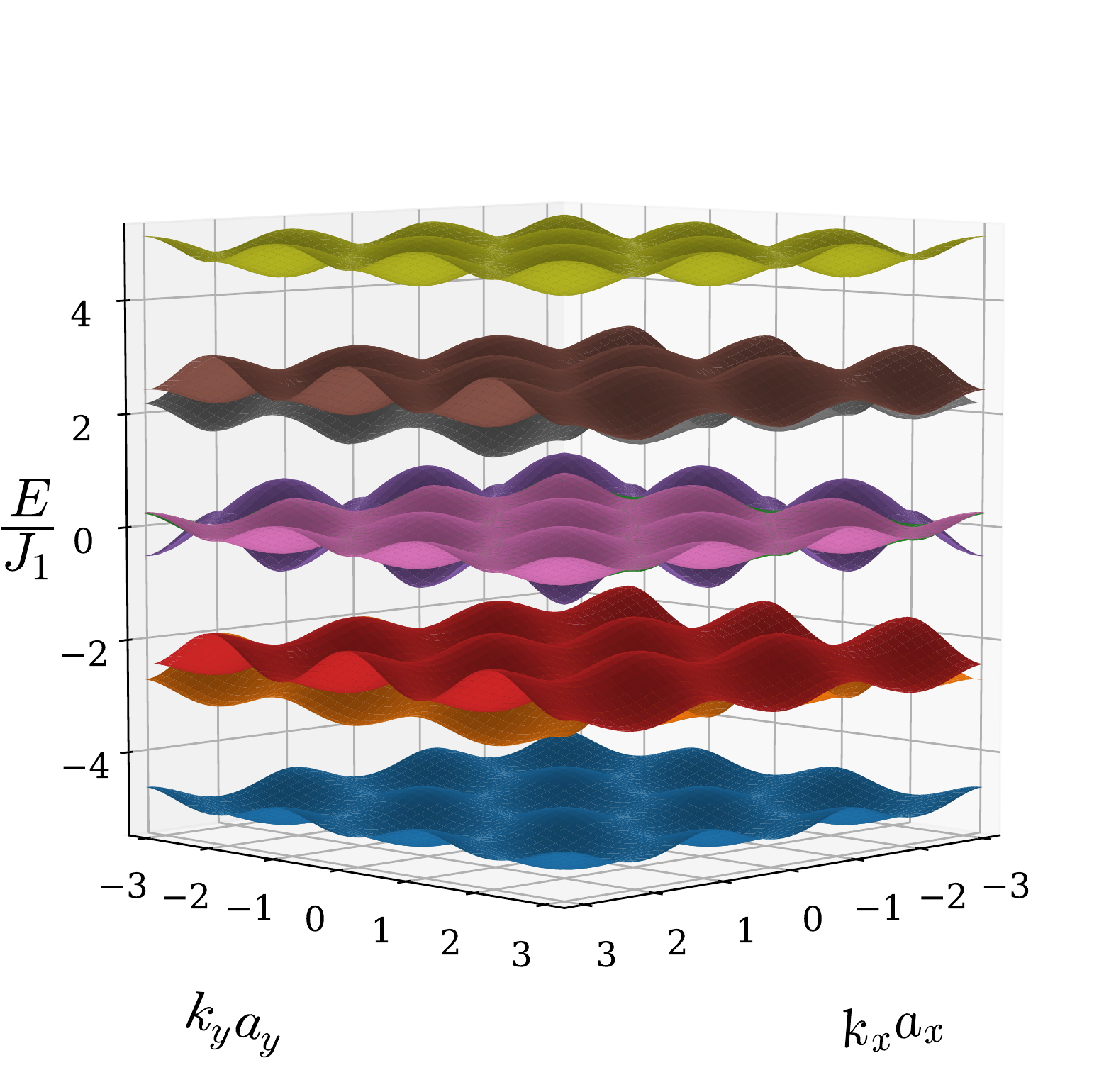}
    \end{center}
    \setlength{\belowcaptionskip}{-10pt}
    \caption{
    Energy spectrum of $\hat{H}_{3d}$ with $k_z a_z = \pi / 3$ and $\phi(t) = \pi / 2$ where $a_x,a_y,a_z$ are the lattice constants of the system. The parameters of $\hat{H}_{3d}$ were set as follows: $J_1=J_2=\Delta_1=\Delta_2$  and 
    $\alpha_1=\alpha_2= 1/3$. }
    \label{fig:4D_Energy_spectrum_for_kz_is_pi_over_3_and_kw_is_pi_over_2}
\end{figure}

We define our 3d time-dependant decoupled Hamiltonian in the following way
\begin{equation}\label{decoupled}
    \hat{H}_{3d}(t) = \hat{H}_1 + \hat{H}_2(t)
\end{equation}
where
\begin{equation}
\begin{aligned}
    \hat{H}_1 = - \sum_{n,m,l} \Big( \Delta_1 & e^{-2\pi\alpha_1 n} \ket{n,m+1,l}\bra{n,m,l} \\
    &+ J_1\ket{n+1,m,l}\bra{n,m,l} + {\rm h.c.} \Big) 
\end{aligned}
\end{equation}
and
\begin{equation}
\begin{aligned}
    \hat{H}_2(t) = -\sum_{n,m,l} \Big( \Delta_2 & e^{i (2\pi \alpha_2 l - \phi(t))} \ket{n,m,l}\bra{n,m,l}\\
    &+ J_2 \ket{n,m,l+1}\bra{n,m,l} + {\rm h.c.} \Big).
\end{aligned}
\end{equation}
We have used $n,m,l$ to label the lattice sites in the $x,y,z$ directions respectively and the time dependence entering the above is $\phi(t) = \frac{2\pi}{T} t$. 
It can be shown that $\hat{H}_1$ is related to the Harper-Hofstadter model via $\hat{H}_1 = \hat{H}_{\rm HH} \hspace{0.2em} \otimes \hspace{0.2em} \hat{\mathds{1}}_z$, where $\hat{H}_{\rm HH}$ represents the 2d Harper-Hofstadter Hamiltonian and the identity matrix has  dimension equal to the number of lattice sites along the $z$ direction. Similarly, $\hat{H}_2$ is related to the 1d time-dependant Aubry-Andre Hamiltonian via $\hat{H}_2 = \hat{\mathds{1}}_{xy} \hspace{0.2em} \otimes \hspace{0.2em} \hat{H}_{\rm AA}$, where $\hat{H}_{\rm AA}$ is the Aubry-Andre Hamiltonian and the identity matrix has a dimension equal to the number of lattice site in the $xy$-plane. We thus have \begin{equation} \label{decoupled 3D H}
    \hat{H}_{3d} = \hat{H}_{\rm HH} \otimes \hat{\mathds{1}}_z + \hat{\mathds{1}}_{xy} \otimes \hat{H}_{\rm AA},
\end{equation}
with
\begin{align} \notag
    \hat{H}_{\rm HH} = - \sum_{n,m} \Big(& \Delta_1 e^{-i2\pi\alpha_1 n}  \ket{n,m+1}\bra{n,m} \\
    &+ J_1\ket{n+1,m}\bra{n,m} + {\rm h.c.} \Big)
\end{align}
and
\begin{align} \notag
    \hat{H}_{\rm AA} = -\sum_{l} &\Big(2  \Delta_2 \text{cos}\big(  2\pi \alpha_2 l - \phi(t)\big) \ket{l}\bra{l} \\ &+ J_2 \ket{l+1}\bra{l}  + J_2 \ket{l}\bra{l+1} \Big)
\end{align}

For our analysis we will be setting $\alpha_1 = \alpha_2 = 1/3$. This extends the unit cell by 3 sites in both the $x$ direction and the $z$ direction. The energy spectrum for $\hat{H}_{3d}$ with periodic boundary conditions can be seen in  \cref{fig:4D_Energy_spectrum_for_kz_is_pi_over_3_and_kw_is_pi_over_2}.
which shows that the lowest-energy band is separated by an energy gap from the others. This band remains gapped (for the chosen parameters of the model) for all values of $k_z$ and for all values of time $t$.
We therefore will calculate the 3d marker for this band. For the system with open boundary conditions,
we take $\Ph$
that projects into the single-particle states having lowest energy that make up 1/9 of the overall
spectrum.

A convenient feature of Hamiltonians of the decoupled  form \cref{decoupled}
with a single occupied band
is that the second Chern number of the system will be the product of the
first Chern numbers of the reduced Hamiltonians \cite{kraus2013four}. 
This is due to the fact that for decoupled Hamiltonians the Bloch states of the system are separable and as a result  the projector is also separable. 
Letting $\Ph'$ project into the lowest band of $\hat{H}_{\rm HH}$ and $\Ph''$ project into the
lowest band of $\hat{H}_{\rm AA}$, we have $\Phs = \Phs' \otimes \Phs''$. 
With this, it can be worked out that
\begin{equation}
    \epsilon_{\mu \nu \rho \sigma}\mathcal{\hat{F}}_{\mu \nu}\mathcal{\hat{F}}_{\rho \sigma} = 8
    \mathcal{\hat{F}}'_{t k_z} \mathcal{\hat{F}}''_{k_x k_y},
\end{equation}
where $\mathcal{\hat{F}}_{ij}$, $\mathcal{\hat{F}}_{ij}'$, and $\mathcal{\hat{F}}_{ij}''$
are constructed from $\Phs$, $\Phs'$, and $\Phs''$ respectively. 
We further note that since these projectors are rank one, they commute and
$
{\rm Tr}(\mathcal{\hat{F}}'_{k_x k_y} \mathcal{\hat{F}}''_{k_z t}) = 
{\rm Tr}(\mathcal{\hat{F}}'_{k_x k_y} ){\rm Tr}(\mathcal{\hat{F}}''_{k_z t}).
$
It is then easy to see that the second Chern number is given by the product of the first Chern numbers of the reduced Hamiltonians.
With this in mind, we deduce that the second Chern number of the system under consideration is one due to the
first Chern number of the bottom band of $\hat{H}_{\rm AA}$ and $\hat{H}_{\rm HH}$ both being one.
Thus we expect that the 3d marker will change by one over a full time period.

\begin{figure}[t]
    \begin{center}
    \includegraphics[width=0.45\textwidth, trim = {0, 2cm ,0 ,8cm}]{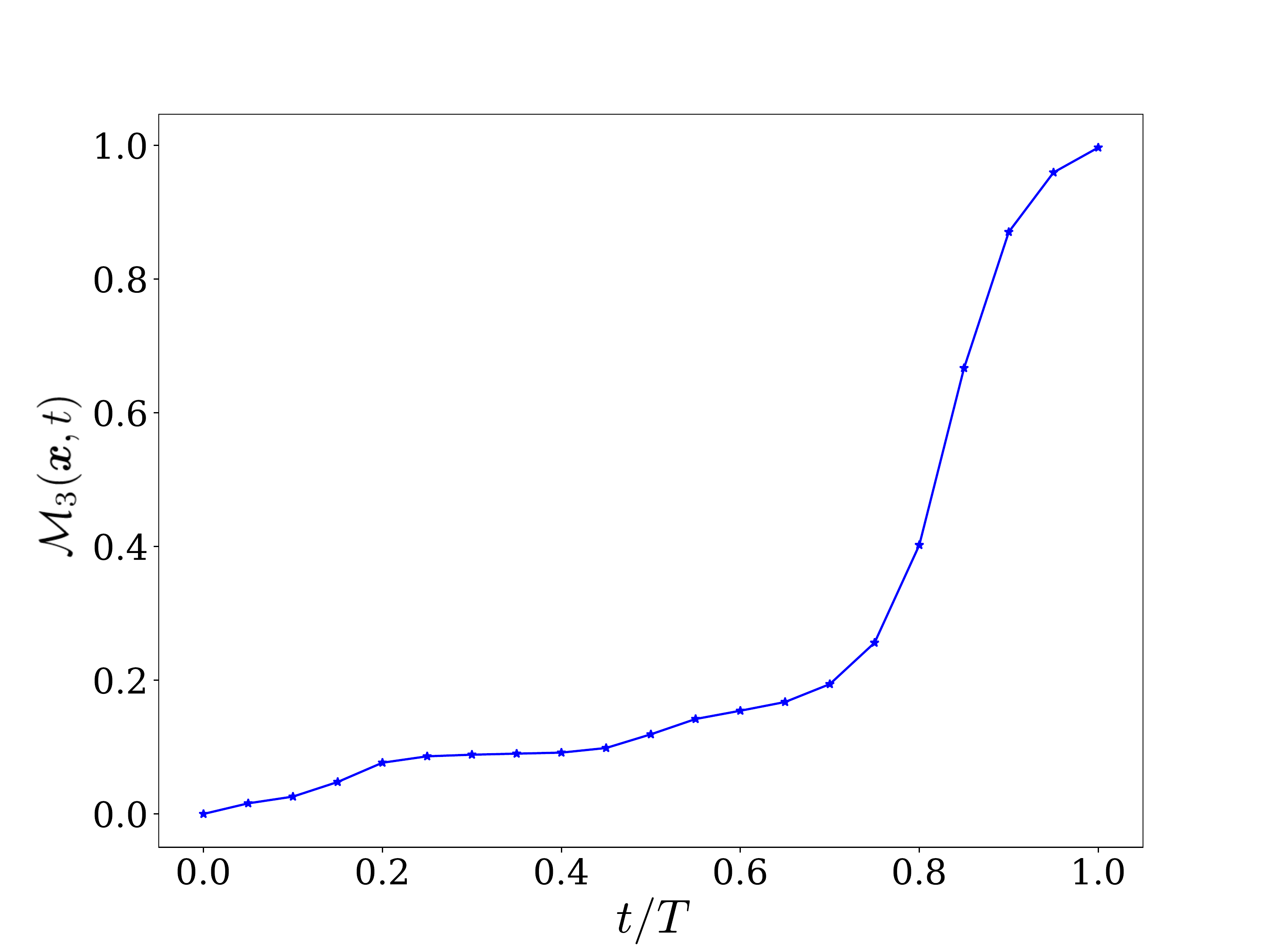}
    \end{center}
    \setlength{\belowcaptionskip}{-10pt}
    \caption{
    The 3d marker ${\cal M}_3(\bx,t)$ as a function of time for $\bx$ deep in the bulk.
    The parameters chosen are the same as those from the previous figure.
    The local trace was taken over the central unit cell of a system containing $10^3$ total
   unit cells. Each cell contains nine sites, and the finite system extends by 10 unit cells in 
   each direction.}
   \label{fig:Change_in_tr_c_PUxUPUyUPUzUP_over_time}
\end{figure}

\cref{fig:Change_in_tr_c_PUxUPUyUPUzUP_over_time} shows the numerical computation of the
 time dependence of ${\cal M}_3(\bx)$
for a finite system with open boundary conditions with $\bx$ deep inside of the bulk. 
The curve was found to be unaffected when a slightly smaller system was taken.
This indeed shows a change of one over the full time period confirming our expectations.
It might also be noted that the curve in \cref{fig:Change_in_tr_c_PUxUPUyUPUzUP_over_time}
shows at least qualitative similarities with the curve in 
 \cref{fig:change in theta_1}. This is not coincidental, but instead is due to the fact that
 we are working with a decoupled model. Indeed, using the decoupled nature of the model, and
 taking the fully translationally invariant system, one can show that the 3d marker is $C_1 \Delta P_1(t)$
 where $C_1$ is the first Chern number of the bottom band of $\hat{H}_{\rm HH}$ and $\Delta P_1(t)$
 is the polarization change of $\hat{H}_{\rm AA}(t)$.

As for the 1d marker, the agreement between the 3d marker and the second Chern number deteriorates when the local trace is take over a unit cell close to the boundary of the system. This is due to the fact that for a unit cell near the boundary of the system, the assumption of translational invariance breaks down. The sum of the 3d marker
over the full system in fact will be periodic in time. It can be worked out that
\begin{align}
\sum_{\bx} {\cal M}_3 (\bx) = \frac{-2i\pi}{3V_c} \epsilon_{ijl} {\rm Tr}(\Xh_i \Xh_j \Xh_l - \Xih_i \Xih_j \Xih_l).
\end{align}
Since the projection operators are periodic in time, this full expression will also be periodic in time.
Also note the direct relation between the above and the extensive axion coupling, $\theta_0$, discussed
earlier in this section.


Before concluding this section, we will say a few words about how the numerical calculations were
carried out.
 Due to the fact that $\hat{H}_{3d}(t)$ is a decoupled Hamiltonian its eigenvectors are separable. As such, to find the eigenvectors of $\hat{H}_{3d}(t)$ we found the eigenvectors of $\hat{H}_{\rm AA}$ and $\hat{H}_{\rm HH}$ and took the tensor product between them. We then constructed $\Pih$ and $\Ph$ and used the same steps outlined for $\mathcal{M}_1({x})$ to calculate $\Uh$. With these we then calculated $\mathcal{M}_3({\bx})$ using \cref{M3} and the subsequent definitions of $\Ah_i$, $\Fh_{ij}$ and $\Fih_{ij}$ in terms of $\Ph$ and $\Uh$.

\section{Conclusion and Future Directions} \label{Section: Discussion and conclusion}

In this paper, we have introduced local topological markers in one and three dimensions
and showed how they can describe topological pumping reflected by the first and second
Chern numbers of the system respectively.  We have motivated and verified the markers both
with general analytical arguments and concrete models. 
Connections between the odd-dimensional markers and Chern Simons forms were explained.
As a secondary result, we have 
generalized the conventional Chern marker to 4d systems. 

Though our expressions are general and can be applied to  systems with multiple occupied bands,
the examples we have focused on involved a single occupied band. For the 1d case, the extension
to multiply occupied bands will be trivial due to band additivity conditions (e.g.\ the first Chern number
for two bands is the sum of the first Chern numbers of the bands treated separately). The 3d case
is considerably richer as there is not such a simple rule for combining second Chern numbers.
As such, a natural future direction would be to investigate multiple-band systems with ${\cal M}_3(\bx)$,
and use its desirable properties (i.e.\ its gauge invariance, locality, and the fact that topological obstructions
do not impede its evaluation) to elucidate the interplay between bands. 

A key feature of local topological markers is the fact that they do not rely on translational
invariance. As such, and as alluded to at times in this paper, there will be multiple interesting systems to consider that possess disorder, both in 1d and 3d. The topological marker can be used to understand how this disorder affects topological 
pumping or, more generally, topological phase transitions.  

Throughout this paper, we have focused on systems with periodic time dependence and
have used an effective adiabatic Hamiltonian \cref{adiabatich} to incorporate dynamics. 
Our expressions for the markers can still be evaluated, however, for arbitrary time dependence
with 
$\Uh$ determined directly from the original microscopic Hamiltonian. 
Investigation of markers modified in such a way is worthwhile.
In \cite{caio2019topological}  it was shown that 
following a quench, the 2d Chern marker undergoes
dynamics resulting in a topological marker current that follows non-trivial scaling
behavior. It will be interesting to investigate similar topological dynamics following a quench 
in 1d and 3d systems
using the markers from this paper and perhaps establish a relation to the 2d case via dimensional
reduction.

Another avenue of investigation would be to analyze the eigenfunctions of the 3d projected antisymmetric moment $\epsilon_{ijl} \Xh_i \Xh_j \Xh_l$. This is motivated by the following. In 1d, the eigenfunctions
of $\Ph \xh \Ph$ are Wannier states. Wannier states have provided a very natural
way of understanding a number of phenomena is solid-state physics 
\cite{RevModPhys.84.1419}. However, their extension from 1d to 2d and 3d can encounter 
insurmountable
difficulties, namely
 exponentially localized Wannier states cannot be constructed when topological obstructions are present.
The relation between the possibility of constructing exponentially localized Wannier states and band 
topology is fundamental and of continued interest \cite{bradlyn2017topological,po2018fragile,bouhon2019wilson}.

On the other hand, one might argue that a possibly natural generalization of $\Ph \xh \Ph$ to 3d is the projected
antisymmetric moment. This is motivated in part by it forming a key ingredient of ${\cal M}_3$ as well
as previous studies where the full trace of this operator was shown to reveal important information \cite{malashevich2010theory,olsen2017surface}. Restricting to the case of a single band, we seek to 
find a linear combination of Bloch states $\ket{W_\ell} = \int_{\rm BZ} d\bk \, \alpha_{\bk \ell} \, \ket{\psi_{\bk}}$
that are eigenstates of $-i \epsilon_{ijl}  \Xh_i \Xh_j \Xh_l$. Using \cref{niu} and considering the
translationally invariant case, this eigenvalue equation
can be written as a first order differential equation in the following way
\begin{align} \label{genWannier}
\epsilon_{ijl} \, (\partial_i \tilde{{\cal A}}_j) \, (i \partial_l + \tilde{{\cal A}}_l) \, \alpha_{\bk \ell} = \lambda_{\ell} \, \alpha_{\bk \ell}
\end{align}
where $\lambda_\ell$ is the eigenvalue.
For the particular model we considered in \cref{Subsection: 3D topological model}, due to the
fact that it is decoupled, the eigenstates are readily found and turn out
to be  
hybrid Wannier functions: $\ket{W_\ell(k_x,k_y)}=\int dk_z e^{-i \ell k_z} \ket{\psi_\bk}$
where the phases of the Bloch states are fixed by \cref{genWannier}.  As indicated earlier, 
the positions of the hybrid Wannier centers will reflect the pumping in the model considered in this paper as
well as other models \cite{olsen2017surface,PhysRevB.101.155130}. In general, however,  
the Berry curvature vector
${\bf \Omega}$ where $\Omega_i=\epsilon_{ijl} \partial_j \tilde{\cal A}_l$ will not be unidirectional. For such cases,
\cref{genWannier} is richer and its solutions will involve integrating along closed streamlines of ${\bf \Omega}$ in
the Brillouin zone. It is very interesting to consider what topological information is contained
in Wannier functions that are generalized in this way.

We hope that the directions specified above will promote  further investigation and
application of local  topological markers. 
\\
\\
After the completion of this work, we became aware of \cite{hayward2020effect} where a 1d topological marker is derived and used to investigate disordered pumping. The marker used in there is similar to the time integral of ${\cal M}_1$ from the present work.

\begin{acknowledgements}
We would like to thank 
Joe Bhaseen, 
Thivan Gunawardana,
Derek Lee, and
Peru d'Ornellas
for particularly useful discussions.  We gratefully acknowledge financial support from EPSRC Grant EP/N509486/1
and a Cecilia Tanner Research Impulse Grant from Imperial College Dept of Mathematics.
\end{acknowledgements}

\appendix

\section{Adiabatic evolution: additional details}
\label{aevolution}
This Appendix is to supplement  \cref{subsection: Adiabatic evolution} from the main text. 
For simplicity, throughout this Appendix we will restrict to the translationally invariant case.
Let the Bloch states of the original 
Hamiltonian be given by $\ket{\psi_{\bk n}} = \ket{\bk} \otimes \ket{\tilde{u}_{\bk n}}$. Here, 
$\ket{\tilde{u}_{\bk n}}$ are eigenstates of the Bloch Hamiltonian of the system.
The projection operators can be constructed as usual as
$\Phs_{\bk} = \sum_n^{\rm occ}  \ket{\tilde{u}_{\bk n}}\bra{\tilde{u}_{\bk n}}$ and
$\Qhs_{\bk} = \sum_n^{\rm unocc}  \ket{\tilde{u}_{\bk n}}\bra{\tilde{u}_{\bk n}}$.

Now define a collection of states within the occupied space as
$\ket{u_{\bk n} }=\sum_{n'}^{\rm occ}  V_{n'n} \ket{\tilde{u}_{\bk n'}}$
where $V_{n'n}$ are elements of a unitary matrix. It is a straightforward exercise to then
verify that $\Phs_{\bk}$ is unchanged when $\ket{\tilde{u}_{\bk n}}$ is
replaced with $\ket{u_{\bk n}}$. 
Our goal is to use this symmetry to bring the adiabatic evolution Hamiltonian to a simple form.
We pick a particular $V$ such that
$\bra{u_{\bk n}} \partial_t u_{\bk n'} \rangle=0$ for any $n$ and $n'$
corresponding to occupied states. Such a unitary matrix can be found from
the matrix differential equation
\begin{align}
i \partial_t V_{nn'} = -\sum_{n''}^{\rm occ}  C_{nn''} V_{n'' n}
\end{align}
with $V_{nn'}(t=0) = \delta_{nn'}$ and $C_{nn''} = i \bra{\tilde{u}_{\bk n}} \partial_t \tilde{u}_{\bk n''} \rangle$.
In practice this equation does not need to be solved. We only need to know
that such a unitary matrix exists.

Next, we perform an identical procedure to find
$\ket{u_{\bk n}}$ for unoccupied bands. A short manipulation then gives
\begin{align}
\hhs_\bk = i{ [ \dot{\Phs}_\bk, \Phs_\bk] }= i \dot{\Phs}_\bk \Phs_\bk + i \dot{\Qhs}_\bk \Qhs_\bk =
i \sum_{n} \ket{\partial_t u_{\bk n}} \bra{u_{\bk n}}.
\end{align}
From this one finds the explicit expression for the time evolution operator from the main text:
\begin{align}
\Uhs_{\bk}(t) = \sum_n \ket{u_{\bk n}(t)} \bra{u_{\bk n}(0)}.
\end{align}

The above considerations are not important when working with a system having a single occupied band.
For the case of multiple occupied bands, it is important to note that $\theta_{\rm CS}$ is gauge invariant
modulo $2\pi$. Thus evaluating $\theta_{\rm CS}$ with the transformed states $\ket{u_{\bk n}}$ will not alter its value up to integer multiples
of $2\pi$ even though these states may not correspond to eigenstates of the original Hamiltonian. The same conclusion applies to the polarization.

\section{Derivation of some key relations}
\label{Appendix:PxQ derivation}
In this Appendix we will provide a derivation of 
\cref{PxQ} from the main text. We also analyse the commutation relation between the projected position operator, $\hat{X}_i$, and an arbitrary translationally invariant operator. 
Making use of \cref{projector} one finds
\begin{align} \notag
    \Ph \xh_i \Qh &= 
    \sum_{\substack{ n \in {\rm occ} \\ m \in {\rm unocc}}}
    \int_{\rm BZ} d \bk d \bk'
     \ket{\psi_{\mathbf{k}n}}\bra{\psi_{\mathbf{k}n}} \xh_i \ket{\psi_{\mathbf{k}'m}}\bra{\psi_{\mathbf{k}'m}} \\
     &= \label{A1}
    \int_{\rm BZ} d \bk d \bk' \ket{\bk} \bra{\bk'}\otimes \Phs_{\bk} \Qhs_{\bk'}  i \partial_{k_i} \delta(\bk - \bk')
\end{align}
where we have used $\bra{\bk} \xh_i \ket{\bk'} = i \partial_{k_i} \delta(\bk - \bk')$.
Next, using that $\Phs_{\bk} \Qhs_{\bk} = 0$, after an integration by parts one finds
\begin{align}
\Ph \xh_i \Qh &= 
     -i     \int_{\rm BZ} d \bk \ket{\bk}\bra{\bk} \otimes  (\partial_i \Phs_{\bk}) \Qhs_{\bk}  
\end{align}
where the delta function has been used to remove one of the integrals. From the relation
$\partial_i \Phs = \partial_i( \Phs^2)$ one can see that $\Phs \partial_i \Phs = \partial_i \Phs \Qhs$.
Using this, one has
\begin{align}
\Ph \xh_i \Qh &= 
     -i  \int_{\rm BZ} d \bk \ket{\bk}\bra{\bk} \otimes  \Phs_{\bk}  \partial_i \Phs_{\bk}
\end{align}
which is the result we are seeking. Taking the adjoint of this equation yields
another useful relation
\begin{align}
\Qh \xh_i \Ph &= 
     i  \int_{\rm BZ}
      d \bk \ket{\bk}\bra{\bk} \otimes    (\partial_i \Phs_{\bk})\Phs_{\bk}.
\end{align}
It is worth noting that these results can be arrived at in another manner by utilizing
a known expression for the matrix element of the position operator
 \cite{blount1962formalisms, sundaram1999wave}
\begin{equation} \label{niu}
    \bra{\psi_{\mathbf{k}n}} \xh_i \ket{\psi_{\mathbf{k}'m}} = (i\delta_{nm}  \partial_i  + (\tilde{\cal A}_{\bk i})_{nm})  \delta(\mathbf{k}-\mathbf{k}').
\end{equation}
One can insert this into the first line of \cref{A1} and proceed from there.

Next, we consider the commutator of $\Xh_i = \Ph \xh_i \Ph$ and an arbitrary translationally invariant
operator $\hat{O} = \int_{\rm BZ} d \bk \; \ket{\bk} \bra{\bk} \otimes \hat{\cal O}_\bk$ where $\hat{\cal O}_\bk$ is
Brillouin zone periodic. 
We further require that
this operator acts only within the projected subspace: $\hat{O} = \Ph \hat{O} \Ph$.
The relation $\xh_i \ket{\bk} = -i \partial_i \ket{\bk}$ can be used to find (in the bulk)
\begin{equation}
\begin{aligned}
[\Xh_i, \hat{O}] &= \Ph \left( \int_{\rm BZ} d\bk (-i) \partial_i ( \ket{\bk}\bra{\bk})  \otimes \hat{\cal O}_\bk \right) \Ph \\
&= \Ph \left( \int_{\rm BZ} d\bk ( \ket{\bk}\bra{\bk})  \otimes  i \partial_i \hat{\cal O}_\bk \right) \Ph \\
&= \int_{\rm BZ} d\bk ( \ket{\bk}\bra{\bk})  \otimes  i \Phs_\bk \partial_i \hat{\cal O}_\bk \Phs_\bk.
\end{aligned}
\end{equation}
Taking the local trace of this expression, we have
\begin{equation}
\begin{aligned}
{\rm tr}_{\bx} \left([\Xh_i, \hat{O}] \right) & = \frac{1}{V_{\rm BZ}} \int_{\rm BZ} d\bk {\rm Tr} 
(\Phs_\bk  \partial_i \hat{\cal O}_\bk \Phs_\bk) \\
& = \frac{1}{V_{\rm BZ}} \int_{\rm BZ} d\bk  \; \partial_i {\rm Tr} 
( \hat{\cal O}_\bk ) = 0
\end{aligned}
\end{equation}
where $\Phs_\bk \hat{\cal O}_\bk \Phs_\bk=\hat{\cal O}_\bk$ and the periodicity of 
$\hat{\cal O}_\bk$  (as a function of $\bk$) have been used.

\section{Relation between $\Ah_i$ and $\Fh_{ij}$}
\label{Appendix:UFU derivation}
It is well known that the conventional Berry curvature and connection are related by
\begin{align}
\tilde{\mathcal{F}}_{ij} = \partial_i \tilde{\mathcal{A}}_j -\partial_j \tilde{\mathcal{A}}_i - i [\tilde{\mathcal{A}}_i, \tilde{\mathcal{A}}_j].
\end{align}
In this Appendix we will seek an analogous relation for the operator versions of these quantities.  Considering the definitions of these
quantities: $\Fh_{ij}=-i [\Xh_i, \Xh_j]$ and $\Ah_i = \hat{U}^\dagger \Xh_i \Uh-\Xih_i$,
it is perhaps not so obvious that there will be an analogous simple relationship.

Let us start with $\hat{U}^\dagger \Fh_{ij} \hat{U}$. Using the defining relation for $\Ah_i$,
this can be written as
\begin{align} \label{B2}
\hat{U}^\dagger \Fh_{ij} \hat{U} &= -i [ \Ah_i + \Xih_i, \Ah_j + \Xih_j] \\
&= \Fih_{ij} -i [ \Ah_i, \Ah_j] -i [ \Xih_i, \Ah_j]+i [  \Xih_j, \Ah_i ] \notag.
\end{align}
Now let us restrict to the translationally invariant case. Using methods very similar to
those in the second half of Appendix \ref{Appendix:PxQ derivation}, one finds
\begin{align}
-i [ \Xih_i, \Ah_j] = \int_{\rm BZ} d\bk \ket{\bk}\bra{\bk} \otimes \Pis_\bk (\partial_i \Ahs_j )\Pis_\bk.
\end{align}
Next, using the translationally invariant expressions for the other quantities in \cref{B2}, e.g.\ 
$\hat{U} =  \int_{\rm BZ} d\bk \ket{\bk}\bra{\bk} \otimes  \Uhs_\bk$, one immediately finds
\begin{equation}
\Uhs^\dagger \Fhs_{ij} \Uhs = {\cal \Pih} (\partial_i \Ahs_j - \partial_j \Ahs_i -i [\Ahs_i,\Ahs_j])   
{\cal \Pih }
+ \Fis_{ij}.
\end{equation}

\providecommand{\noopsort}[1]{}\providecommand{\singleletter}[1]{#1}%

\end{document}